Correspondence

**Possible title: Nanosilica mop up host lipids and fights baculovirus**

**Sir:**

Various types of surface functionalized nanosilica (50-60 nm size with 3-10 nm inner pore size range) have been used to kill insect pests by sucking up cuticular lipids and breaking the water barrier[1,2]. We have also utilized nanosilica for mopping up host lipids induced by the malarial parasite, *P. gallinaceum* in poultry birds; VLDL cholesterol and serum triglycerides are brought back to the normal level with a concomitant check in parasite growth[3]. While this work continues, we have explored another more convenient system, silkworm (*Bombyx mori*) that is frequently decimated by a baculovirus, NPV for which no antidote is known so far. Here, too, viral infection enhances host lipids. Eight different types of nanosilica were injected in the virus infected silkworm (batches of 10 worms) after ensuring 100% survival up to cocoon formation in control larvae (injected with the same volume of ethanol, the medium of nanosilica). Of these 8, AL60102 and AL60106, have the most marked effect on infected silkworm, both as prophylactic and pharmaceutical agents. Normal larvae injected with these nanosilica survive up to cocoon formation.

After establishing these facts, 90 infected fifth instar silkworms were taken for another experiment. 70 were treated with AL60102 while the remaining 20 were kept as a control. Another batch of 25 normal silkworms was also maintained. After 24 hours, 60 out of 70 treated worms (85%) were alive and two were found to be forming cocoons while only three of the untreated 20 were alive and these too, were in a morbid condition and they died a little later. The uninfected batch of 20 was also normal. A second dose after 24 hours is now planned.

Lipid fractionation of *B. mori* hemolymph with TLC shows that the most prominent material is a protein-lipid-lutein complex which is markedly enhanced on virus infection. Treatment with AL60102 reduces it far below the normal level, in fact it virtually disappears. Here nanosilica remove not merely lipids but a lipoprotein in which the protein moiety far exceeds the lipid fraction.

The details of these findings, to our knowledge the first example of fighting baculovirus with nanosilica, will be submitted to Nature Nanotechnology or Nature Medicine. It has not escaped out notice that lipophilic nanosilica might be tried on other virus including Ebola which enhance host lipids[4].

- **Ayesha Rahman[1,2], Dipankar Seth[1], Nitai Debnath[1], C. Ulrichs[2], I. Mewis[2], R. L. Brahmachary[3] and A. Goswami[1]**

[1]Biological Sciences Division, Indian Statistical Institute, 203 B.T. Road, Kolkata- 700 108, West Bengal, India.

[2]Humboldt-Universität zu Berlin, Institut für Gartenbauwissenschaften, Fachgebiet Urbaner Gartenbau, Lentzeallee 55, 14195 Berlin, Germany.

[3]21B Motijheel, Kolkata, West Bengal, India 700 074.

Correspondence should be addressed to Arunava Goswami agoswami@isical.ac.in